\documentclass[letterpaper, 10 pt, conference]{ieeeconf}  % Comment this line out if you need a4paper

\IEEEoverridecommandlockouts                              % This command is only needed if 
\newcommand{\tabincell}[2]{\begin{tabular}{@{}#1@{}}#2\end{tabular}} %%table
\usepackage{cite}
\usepackage{textcomp}
\usepackage{threeparttable}
\usepackage{multirow}
\usepackage{graphicx}
\usepackage{booktabs} % For formal tables
\usepackage{float}
\usepackage{color}
\usepackage[noend]{algpseudocode}
\usepackage{algorithmicx,algorithm}
\usepackage{times}
\usepackage{epsfig}
\usepackage{enumerate}
\usepackage{verbatim}
\usepackage{cite}
\usepackage{amsfonts}
\usepackage{array}
\usepackage{comment}
\usepackage{multicol}
\usepackage{multirow}
\usepackage{epsfig}
\usepackage{booktabs}
\usepackage{float}
\usepackage{graphicx}
\usepackage{amsmath,bm}
\usepackage{amssymb}
\usepackage{amsfonts}
\usepackage{graphicx,lipsum}
\usepackage{url}
\title{\LARGE \bf
Instrumental Variable Learning for Chest X-ray Classification
}

\author{Weizhi Nie$^{1,2}$, Chen Zhang$^{1}$, Dan song$^{1*}$, Yunpeng Bai$^{3}$, Keliang Xie$^{4}$, and Anan Liu$^{1}$% <-this % stops a space
\thanks{*This work was supported by the National Natural Science Foundation of China (62272337)}% <-this % stops a space
\thanks{$^{1,2}$Weizhi Nie is with the School of Electrical and Information Engineering, Tianjin University, Tianjin 300072, China, and Institute of Artificial Intelligence, Hefei Comprehensive National Science Center, Hefei 230088, China}
\thanks{$^{1}$Chen Zhang, $^{1}$Dan Song, and $^{1}$Anan Liu are with the School of Electrical and Information Engineering, Tianjin University, Tianjin 300072, China. {\tt\small *Corresponding author: dan.song@tju.edu.cn}}
\thanks{$^{3}$Yunpeng Bai is with the Department of Cardiac Surgery, Chest Hospital, Tianjin University, and Clinical school of Thoracic, Tianjin Medical University, Tianjin 300052, China. {\tt\small oliverwhite@126.com}}
\thanks{$^{4}$Keliang Xie is with the Department of Critical Care Medicine, Department of Anesthesiology, and Tianjin Institute of Anesthesiology, Tianjin Medical University General Hospital, Tianjin 300052, China. {\tt\small xiekeliang2009@hotmail.com}}
}

\begin{document}

\maketitle
\thispagestyle{empty}
\pagestyle{empty}

%%%%%%%%%%%%%%%%%%%%%%%%%%%%%%%%%%%%%%%%%%%%%%%%%%%%%%%%%%%%%%%%%%%%%%%%%%%%%%%%
\begin{abstract}

The chest X-ray (CXR) is commonly employed to diagnose thoracic illnesses, but the challenge of achieving accurate automatic diagnosis through this method persists due to the complex relationship between pathology. In recent years, various deep learning-based approaches have been suggested to tackle this problem but confounding factors such as image resolution or noise problems often damage model performance. In this paper, we focus on the chest X-ray classification task and proposed an interpretable instrumental variable (IV) learning framework, to eliminate the spurious association and obtain accurate causal representation. 
Specifically, we first construct a structural causal model (SCM) for our task and learn the confounders and the preliminary representations of IV, we then leverage electronic health record (EHR) as auxiliary information and we fuse the above feature with our transformer-based semantic fusion module, so the IV has the medical semantic. Meanwhile, the reliability of IV is further guaranteed via the constraints of mutual information between related causal variables. 
Finally, our approach's performance is demonstrated using the MIMIC-CXR, NIH ChestX-ray 14, and CheXpert datasets, and we achieve competitive results. 

\end{abstract}

\begin{keywords}
Instrumental Variable, Causal Inference, Medical Image Processing, Chest X-ray Images Classification.
\end{keywords}

%%%%%%%%%%%%%%%%%%%%%%%%%%%%%%%%%%%%%%%%%%%%%%%%%%%%%%%%%%%%%%%%%%%%%%%%%%%%%%%%
\section{INTRODUCTION}

The chest X-ray (CXR) is frequently utilized for early screening of various diseases, including heart disease and thoracic diseases, as it is a non-invasive test and is easy of accessibility and lower cost. However, in clinical practice, medical image analysis is a critical task that requires significant time and effort from medical professionals, especially when it comes to the manual interpretation of chest X-rays, and the accuracy of the results may be influenced by individual medical expertise \cite{brady2012discrepancy}. Recent advances in deep learning have shown promise in assisting with this process \cite{mao2022imagegcn,ouyang2020learning}, by leveraging convolutional neural networks and other deep learning techniques, it is possible to automatically detect and classify various pathologies present in chest X-rays, including pneumonia and tuberculosis. This can help to reduce the workload of radiologists and improve the accuracy and efficiency of diagnosis. 
However, with the increasing use of deep learning methods, researchers have found that deep learning models often lack interpretability due to their black-box nature. However, with the development of causal inference theory \cite{glymour2016causal,pearl2000models,pearl2014interpretation}, it has become clear that this theory can be leveraged to improve the interpretability of deep learning models. Specifically, in the context of optimizing the accuracy of deep learning-based classification systems, researchers have begun to explore the use of causal inference to enhance the interpretability of these models \cite{sui2022causal}. By incorporating causal inference techniques into deep learning models, researchers hope to not only improve the accuracy of chest X-ray classification, but also make these models more transparent and interpretable, which could be crucial for clinical decision-making and patient safety \cite{ribeiro2023learning}, and there is hope that the spurious correlations between pathology or inherent problem of data caused by confounders \cite{pearl2009causal} can be solved.

\begin{figure}
\centering
\includegraphics[width=0.95\linewidth]{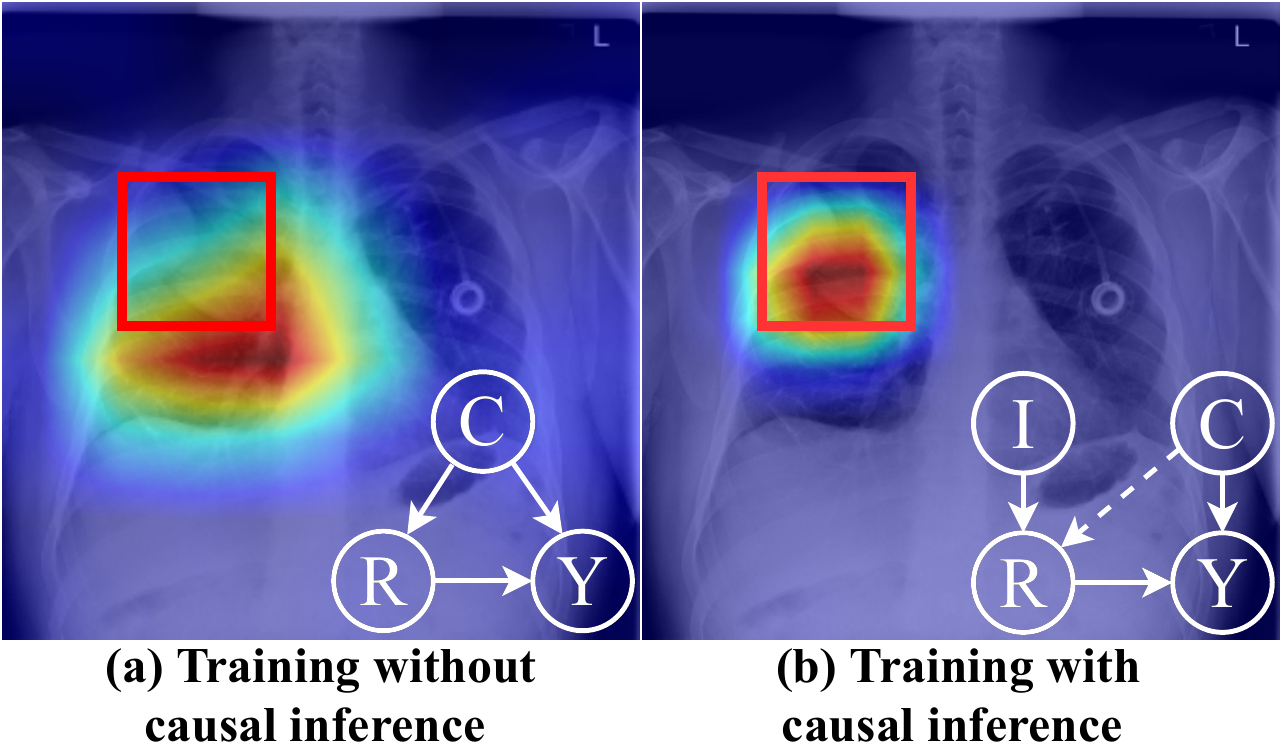}
\caption{An example of prediction. The SCM and corresponding visualization result with disease bounding box of (a) traditional deep learning-based method training without causal inference and (b) causal inference method. ``C'', ``R'', ``Y'', and ``I'' denote confounding, causal feature representation, prediction result, and instrumental variable, respectively.}
\label{example}
\end{figure}

Based on the causal theory in \cite{pearl2009causal}, if the cause $X$ or the causal representation $R$ and its effect $Y$ have the common cause $C$, we can denote $C$ as the confounding. 
Taking Fig.\ref{example} (a) as an example, CXR images pose certain inherent challenges that are not easy to overcome, including high interclass similarity \cite{rajaraman2020training}, atypical and noisy data, complex co-occurrence patterns between diseases \cite{wang2017chestx}, as well as long-tailed or imbalanced data distribution \cite{zhang2021deep}. Besides, traditional deep learning-based models inevitably suffer from a performance drop in an out-of-distribution (OOD) scenario simply because the probability distributions of the training and test data are different. All of these problems can be called potential confounders $C$, as they affect the feature learning $R$ and prediction results $Y$. For example, we know it is a challenging task to classify some rare categories of CXR images, but if these images have some kind of letter marker or medical device in them just because the same disease may be treated similarly, the deep model tends to tackle the task by aforementioned confounding features. The Gradient-weighted Class Activation Mapping (CAM) \cite{selvaraju2017grad} in (a) shows that the diagnostic effect without causal inference has a certain deviation from the true label. Thus, how can we eliminate the confounding effect?

\begin{figure}
\centering
\includegraphics[width=0.75\linewidth]{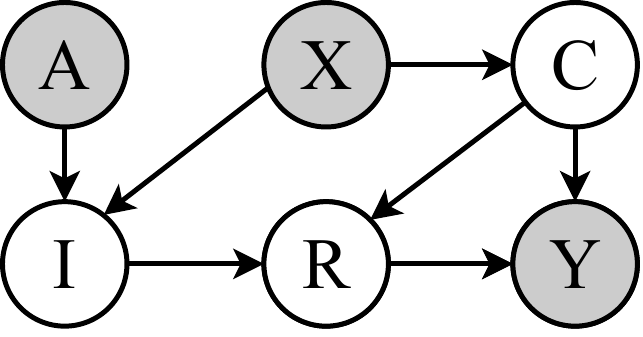}
\caption{The SCM of our proposed framework. ``A'', ``X'', ``C'', ``R'', ``Y'', and ``I'' denotes auxiliary information, CXR data, confounders, causal feature representation, prediction result, and instrumental variable, respectively.}
\label{scm}
\end{figure}

According to the causal theory by Pearl \cite{glymour2016causal}, we can use some methods to cut the side effect of the confounding $C$, then the deep model can locate which pixel is the true cause of the disease. As shown in Fig.\ref{example} (b), we can find the disease classification performance is better than the traditional convolution-based model without causal settings from the visualization result, which is closer and more precise to the true label. In our work, we see the CXR classification task from a causal perspective, which limits the adverse effects of confounding factors on model performance and improves the interpretability and stability of our proposed model. There are several causal intervention methods such as backdoor adjustment, and front-door adjustment, these two methods need all confounders $C$ to be observed, which is hard to satisfy the condition as the complicated nature of the medical scenario, so the related method using backdoor or front-door adjustment always ease the constraint by making some assumptions or approximate \cite{yue2020interventional}. Fortunately, there is another way to make a causal intervention, named instrumental variables, which require no above constraint. From the causal theory by Pearl in \cite{pearl2000models}, the structural causal model (SCM) \cite{glymour2016causal} shown in Fig.\ref{example} (b) defines the fashion of instrumental variable, the nodes $I$, $C$, $R$, $Y$ represent variables of instrumental variables, confounders, causal representation, and prediction results, respectively. A valid IV can eliminate the confounding effect according to Pearl \cite{pearl2000models}. In this work, we aim to CXR classification modeling with CXR data, as the SCM showed in Fig.\ref{scm}. The key contributions of our work are followed as:
\begin{itemize}
\item We propose a novel instrumental variable learning framework that can address the abnormality classification problem in Chest X-rays. This framework can eliminate the confounding effect and make the prediction results understandable and reliable;
\item The framework we proposed contains two parts: an instrumental variable estimation learning module and a medical semantic fusion module. The ablation study proves the effectiveness of the combination of the three parts;
\item We have validated the effectiveness of our method using several popular datasets and compared it to several current efficient methods. Our final experimental findings demonstrate that our method is superior in performance;
\end{itemize}

The remainder of our paper is outlined below. In Section \ref{sec:related work}, we present the related work. The proposed approach is presented in Section \ref{sec:approach}. In Section \ref{sec:experiment}, we present key experiments. This section includes a summary of the experimental results, in which we demonstrate the effectiveness of our approach by using it to solve various classification problems related to thoracic diseases. 
Finally, in Section \ref{sec:conclution}, we discuss and make a conclusion for this work.

\section{RELATED WORK}
\label{sec:related work}

\subsection{CXR Classification with Deep Learning}
Deep learning-based CXR classification methods showcasing amazing performance in recent years \cite{wang2017chestx,ouyang2020learning,rocha2022attention,minaee2020deep,ke2021chextransfer,paul2021discriminative}, the general idea of previous works is to extract the image-level feature by a neural network framework such as ResNet \cite{he2016deep} or DenseNet \cite{huang2017densely}, then feed the feature into a classifier to get the classification score. 
Ouyang \textit{et al.}\cite{ouyang2020learning} trained a hierarchical attention framework to get the hierarchical attention representation step by step, their algorithm includes explicit ordinal attention constraints, enabling principled model training in a weakly-supervised fashion and facilitating the generation of visual-attention-driven model explanations.
Ratul \textit{et al.}\cite{ratul2021multi} proposed a model built on a dense convolutional neural network that operates through two stages. In the first stage, the model is trained solely on images from patients with thoracic diseases. Then, in the second stage, the entire network is trained on the complete dataset, which includes chest X-rays from both healthy and unhealthy patients.
Mao \textit{et al.}\cite{mao2022imagegcn} highlighted the significance of modeling image-level relations in computer vision to improve image representation and model consistency. They thought existing approaches often treat each input image independently, limiting their effectiveness, so the authors proposed a graph convolutional network framework for inductive multi-relational image modeling using both original pixel features and their relationships with other images. 

Previous works mainly focus on the proper design of model structure or loss, though some works tend to model the relationship among CXR images, according to the causal theory we know that correlation is not causation, so they capture correlations at best. Different from their methods, in this study, we aim to explore the true causal relationship between the pathology area and the corresponding disease, and get rid of the side effect of latent confounders.

\subsection{Causal Inference with Deep Learning}
Causal inference \cite{sobel1996introduction} is a powerful methodology that can be applied to numerous domains, including statistics \cite{pearl2009causal} and epidemiology \cite{richiardi2013mediation}. Causal inference aims to pursue causal effects and eliminate false bias \cite{bareinboim2012controlling}. Recently, causal inference is applied in computer vision tasks using deep learning-based methods \cite{wang2020visual,yang2021deconfounded,yue2020interventional,zhang2020causal,li2021causal}. 
Specifically, Yue \textit{et al.}\cite{yue2020interventional} argued that the performance of the few-shot learning model is indeed limited by the confounding factor of pre-trained knowledge, so they use a SCM to model the problem and delete the bias via the backdoor adjustment with some assumption for simple.
Wang \textit{et al.}\cite{wang2020visual} proposed a causal intervention-based framework to solve image captioning and visual question answering tasks, the key idea is ``borrow and put'', i.e. borrowing objects $Z$ from other images and then placing them around the current image, then evaluating whether the presence of $Z$ affects the relationship in the current image. Using $do(\cdot)$ as causal intervention, this work is conducting the tasks by $P(Y|do(X))$ rather than traditional $P(Y|X)$, where $Y$ is the predicting result and $X$ is the potential cause.
Li \textit{et al.}\cite{li2021causal} proposed a causal Markov model based on variational autoencoder (VAE) structure to disentangle disease-related variables. However, the back/front-door path is usually hard to find due to the complicated clinical scenario.

In our work, we leverage a SCM shown in Fig.\ref{scm} to analyze the causalities between entities, different from previous work using back/front-door adjustment to realize causal inference, we leverage instrumental variables, which do the intervention without observed confounders.

\section{APPROACH}
\label{sec:approach}
\begin{figure*}
\centering
\includegraphics[width=0.95\linewidth]{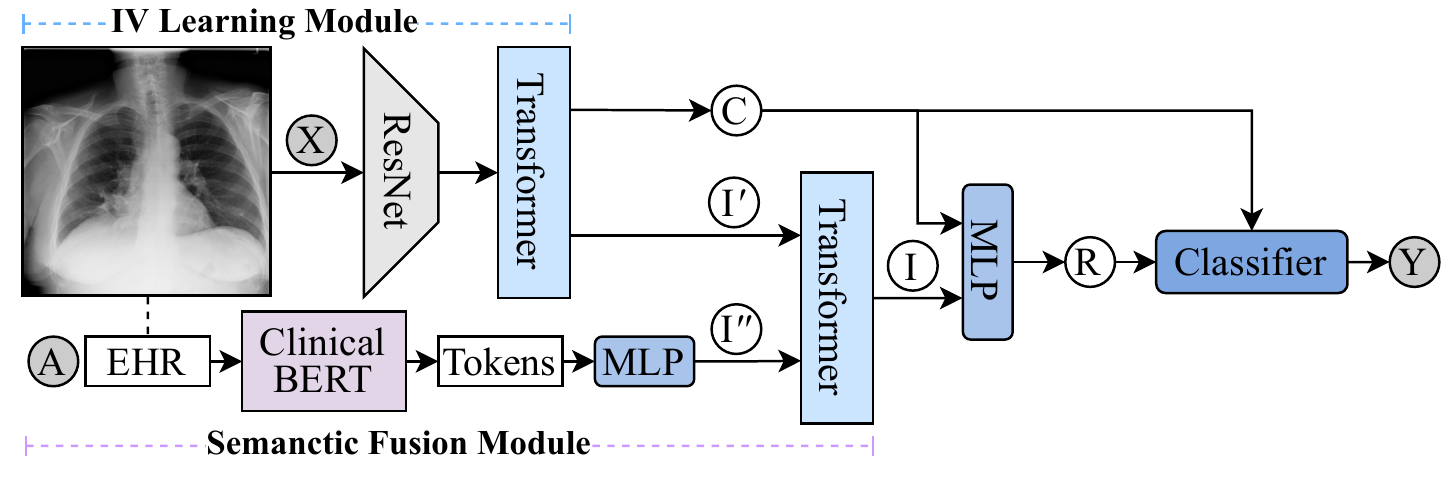}
\caption{Overview of our framework. The whole framework consists of an IV learning module and a semantic fusion module. For the IV learning module, we leverage a transformer to get $C$ and preliminary IV $I^{'}$, then we use another transformer to fuse the feature $I^{''}$ from auxiliary information $A$ to make the IV have medical semantic. After that, $C$ and $I$ are fed into MLP to get causal representation $R$, then we can get the classification scores via a classifier.}
\label{framework}
\end{figure*}

In this paper, we propose an instrumental variable learning method for chest X-ray classification, our framework is shown in Fig.\ref{framework}. It contains two modules: the IV learning module and the medical semantic fusion module. The IV learning module is proposed to learn the representation of confounder and IV, some constraints to make IV valid and meet the two conditions mentioned earlier are applied at the same time. Then, the semantic fusion module is introduced to make the IV representation have medical semantics.

\subsection{Preliminaries} 
The causal intervention aims to eliminate the effect of $C$ on $R$ and $Y$, which is denoted as the $do$ operator mentioned above. A valid IV can block the path of $R \xleftarrow{} C \xrightarrow{} Y$ and implement causal intervention by the colliding junction $I \xrightarrow{} R \xleftarrow{} C$, which entails $C \xrightarrow{} R$ blocked. 
More formally, for a valid IV, the following two conditions must be met: a) $I$ is independent of $C$, i.e. $Corr(I,C) = 0$, where $Corr(\cdot)$ denotes the correlation coefficient function; and b) the IV $I$ affects $Y$ only via $R$, i.e. $Corr(I,R) \neq 0$ and $P(Y|I,R,C) = P(Y|R,C)$. If the above two conditions are met, the IV can make the $C \xrightarrow{} R$ blocked.
Therefore, how to make the IV valid is of vital importance in the causal inference scenario, but the complex clinical factors make it tougher. Inspired by AutoIV \cite{yuan2022auto}, we design a module based on the two constraints for IV to learn a valid IV from other observed variables, furthermore, we can give the learned IV semantics to make it explainable in the clinical scenario. To this end, we construct a SCM for the CXR classification task shown in Fig.\ref{scm}, 
% where the valid IV is learned from the CXR data $X$ and auxiliary information $A$, then we can acquire the causality between the feature representation $R$ and the prediction $Y$.
each arrow in the directed acyclic graph denotes the causality and direct effect, and the solid dots $A, X, Y$ denote variables that can be observed. The IV ($I$) is learned from the CXR data ($X$) and auxiliary medical information ($A$) with medical semantic meaning such as the patients' congenital attribute, the constraints of mutual information makes the IV valid \cite{yuan2022auto}. Then a valid IV can help to eliminate the spurious correlation between $R$ and $Y$. 

\subsection{Instrumental Variable Learning}
As shown in Fig.\ref{framework}, the CXR image $X$ is fed to ResNet \cite{he2016deep} to get the initial feature representation, then we fed it to a Transformer \cite{vaswani2017attention} to get the preliminary decoupling representation of the confounder $(C)$ and the IV $(I)$. 
Specifically, Given a CXR image $x \in \mathbb{R}^{H_0 \times W_0 \times 3}$ as input, we extract its spatial feature $F \in \mathbb{R}^{H \times W \times d}$ using the backbone, where $H_0 \times W_0$, $H \times W$ represent the height and width of the CXR image and the feature map respectively, and $d$ denotes the hidden dimension of the network, where we set $d = 2048$ in this work. 
Then, we adopt zero-initialized $I_0^{'} \in \mathbb{R}^{k \times d}$ as the queries in the cross-attention module inside the transformer, each decoder layer $l$ updates the queries $I^{'}_{l-1}$ from its previous layer, $k$ denotes the disease categories. Here, we denote $I'$ as the IV representation without medical semantics and $C$ as the confounding feature as follows:
\begin{equation}
\begin{aligned}
&I^{'}_l = softmax(I^{'}_{l-1}F/\sqrt{dim_{F}})F,\\
&C_l = (1 - softmax(I^{'}_{l-1}/\sqrt{dim_F}))F,
\end{aligned}
\label{decoder}
\end{equation}
where a such method is not enough to disentangle, we want to impel the model to learn both IV and confounding features via the IV learning module, so we feed them to linear layers and design corresponding loss functions to achieve that:
\begin{equation}
\begin{aligned}
&z_I = Cls_I(Lin_I(I_l)),\mathcal{L}_{I}= -\frac{1}{\lvert X \rvert}\sum_{x \in X} y^\top\log(z_I),\\
&z_C = Cls_C(Lin_C(C_l)),\mathcal{L}_{C}= -\frac{1}{\lvert X \rvert}\sum_{x \in X} KL(y_{u}, z_C)
\end{aligned}
\label{loss1}
\end{equation}
where ``Cls'' denotes the classifier, ``Lin'' represents linear layer, $z$ is the classification logits, $y$ is the ground truth, $y_u$ is a uniform distribution and ``KL'' denotes the KL-Divergence. These two loss functions aim to learn stable IV features and make the confounders less relevant to causal features. Until here, we assume we get the preliminary feature of $I$ and $C$ in Fig.\ref{framework}. 

CXR image $X$ is considered exogenous \cite{yuan2022auto}, then our $I$ and $C$ are also exogenous. Then, we concatenate $I$ and $C$, and feed them into a point-wise multi-layer perceptron (MLP) network to get the causal feature $R$. Finally, we can get the classification result $Y$ by a classifier, which input is the concatenation of $R$ and $C$, i.e. $P(Y|X) = Cls_Y(R\oplus C)$.

\subsection{Semantic Fusion}
As we mentioned in section \ref{dataset}, there are some congenital attributes that can be utilized from the three datasets. We adopt ClinicalBERT \cite{huang2022clinicalbert} and transformer to process the auxiliary information $A$. 
Specifically, we get the EHR word token first, each tokenized EHR is preceded by a classification token \cite{irvin2019chexpert}, while ``EHR'' in NIH and CheXpert datasets are short with only a few words, so we get the text embeddings by zero padding to 256. We assume that the high dimensional feature contains the cause of disease $I^{''}$, then we adopt a multi-layer transformer to fuse $I$ and $I^{''}$ to enforce the IV to have the medical semantic.

\subsection{Constraints of Mutual Information}
\label{constarints}
As we mentioned above, a valid IV should be independent of the confounder, and should affect the result only through causal features. Thus we utilize the mutual information, denote as $M$ to realize a) $E(C|G) = E(C)$ and b) $P(R|I) \neq P(R), P(Y|I,R,C) = P(Y|R,C)$. To achieve the above constraints, we minimize $M(I,C)$, maximize $M(I,R)$, and minimize $M(I,Y)$ at the same time.

Minimizing $M(I,C)$ aims to eliminate the confounding effect \cite{glymour2016causal}. we designed a loss function based on the mutual information:
\begin{equation}
\mathcal{L}_{IC}= \frac{1}{X^2}\sum_{i \in X}\sum_{j \in X}(\log f_{\phi}(C_i|I_i) - \log f_{\phi}(C_j|I_i)),
\label{ic}
\end{equation}
where $f_\phi$ denotes a MLP $f$ with parameters $\phi$, which with the propose of approximating $P(C|I)$, the feature representation $C$ and $I$ are learned via Eq.(\ref{loss1}). Similarly, to maximize $M(I,R)$, we design the following loss:
\begin{equation}
\mathcal{L}_{IR}= -\frac{1}{X^2}\sum_{i \in X}\sum_{j \in X}(\log f^{'}_{\phi^{'}}(R_i|I_i) - \log f^{'}_{\phi^{'}}(R_j|I_i)).
\label{ir}
\end{equation}

Meanwhile, we minimize the mutual information of $P(Y|I,R,C)$ via the following loss:
\begin{equation}
\begin{aligned}
\mathcal{L}_{IY}=&\frac{1}{X^2}\sum_{i \in X}\sum_{j \in X}(\log f^{''}_{\phi^{''}}(Y_i|I_i)\\
&- \log f^{''}_{\phi^{''}}(Y_j|I_i)).
\end{aligned}
\label{iy}
\end{equation}

\subsection{Optimization}
Till now, we have learned the representation of $I$, $C$, and $R$, then we apply a supervised learning loss for the classifier to get the prediction $Y$ to realize $P(Y|X)=Cls_Y(R\oplus C)$:
\begin{equation}
\mathcal{L}_{Y}= -\frac{1}{\lvert X \rvert}\sum_{x \in X} y^\top\log(R\oplus C),
\label{lossy}
\end{equation}

Thus, our training objective is defined as:
\begin{equation}
\mathcal{L}=\mathcal{L}_{I}+\mathcal{L}_{C}+\mathcal{L}_{IC}+\mathcal{L}_{IR}+\mathcal{L}_{IY}+\mathcal{L}_{Y},
\label{total}
\end{equation}
which we aim to learn features representation of variables and make the instrumental variable valid at the same time.

\section{EXPERIMENT}
\label{sec:experiment}

\subsection{Data Sets and Evaluation Metrics}
\label{dataset}
We test our framework for disease classification on three mainstream public data sets, NIH ChestX-ray14 \cite{wang2017chestx}, CheXpert \cite{irvin2019chexpert}, and MIMIC-CXR \cite{johnson2019mimic} dataset.
The area under the ROC curve, namely AUC is measured for evaluation.

\subsubsection{Chest X-ray14}
The NIH ChestX-ray14 dataset contains 112,120 frontal-view X-ray images of 30,805 unique patients. Of these patients, 56.49\% are male and 43.51\% are female. Out of the total number of images, 51,708 have been labeled with up to 14 pathologies.
% , which include ``Cardiomegaly'', ``Pneumothorax'', ``Consolidation'', ``Mass'', ``Pleural Thickening'', ``Infiltration'', ``Edema'', ``Hernia'', ``Fibrosis'', ``Emphysema'', ``Pneumonia'', ``Nodule'', ``Atelectasis'', and ``Effusion''. The remaining 60,412 images are labeled as ``No Finding''. 
Our experiments on this dataset are conducted using the official patient-level data split, with all images randomly shuffled.
This data set provides the patients' individual attributes such as age and gender, which can be adopted as auxiliary information in our framework.

\subsubsection{MIMIC-CXR}
The MIMIC-CXR dataset contains 377,110 images of 65,379 patients with up to 227,835 anonymized electronic health records (EHRs), each EHR is labeled with specific disease class observations, noted as positive, negative, and uncertain. 
We use the patient and view information as auxiliary information in our framework.

\subsubsection{CheXpert}
The CheXpert dataset contains 224,316 CXR images of 65,240 patients, which also provides information such as age and image view.

During the training process, we resize each CXR image in data sets to 256 $\times$ 256 and crop it to 224 $\times$ 224 without any data augmentation methods.

\subsection{Implementation Details}
We use PyTorch for implementation. Our experiment is operated by using NVIDIA GeForce RTX 3090 with 24GB memory. We use the Adam\cite{kingma2014adam} optimizer with True-weight-decay\cite{loshchilov2017decoupled} 1$e$-2 and the maximal learning rate is 1$e$-3.

\subsection{Comparison with State-of-the-art Methods}
\subsubsection{NIH ChestX-ray14}
We first compared with some SOTA methods on the NIH ChestX-ray14 dataset.
The AUC score of each pathology is summarized in Table.\ref{tb1}, and the best performance of each pathology is shown in bold.
Our baseline achieved an average AUC score of 0.867 across 14 thoracic diseases, which is competitive with previous works.

\begin{table}[t]
\setlength\tabcolsep{3pt}
  \centering
  \caption{Comparation Of AUC Scores with previous SOTA Works on NIH ChestX-ray14 dataset}
    \begin{tabular}{c|c|c|c|c|c}
    \toprule
    \tabincell{c}{Abnormality}&\tabincell{c}{Wang \textit{et al.}\\\cite{wang2017chestx}  }&\tabincell{c}{Xi \textit{et al.}\\\cite{ouyang2020learning}}&\tabincell{c}{ImageGCN\\\cite{mao2022imagegcn}}&\tabincell{c}{DGFN\\\cite{gong2021deformable}}&\tabincell{c}{Ours}\\
    \midrule
    Atelectasis & 0.70  & 0.77  & 0.80 & 0.82 & \textbf{0.84} \\
    \midrule
    Cardiomegaly & 0.81   & 0.87  & 0.89 &\textbf{0.93} & 0.88 \\
    \midrule
    Effusion & 0.76   & 0.83  & 0.87 &\textbf{0.88} & 0.87 \\
    \midrule
    Infiltration & 0.66   & 0.71  & 0.70 & 0.75 & \textbf{0.78} \\
    \midrule
    Mass  & 0.69    & 0.83  & 0.84 & \textbf{0.88} & 0.85 \\
    \midrule
    Nodule & 0.67   & 0.79  & 0.77 & 0.79 & \textbf{0.80}  \\
    \midrule
    Pneumonia & 0.66    & 0.82  & 0.72 & 0.78& \textbf{0.89}  \\
    \midrule
    Pneumothorax & 0.80     & 0.88  &0.90  &0.89 & \textbf{0.91} \\
    \midrule
    Consolidation & 0.70    & 0.74  & 0.80 & 0.81 & \textbf{0.86} \\
    \midrule
    Edema & 0.81    & 0.84  & 0.88 &0.89 & \textbf{0.92} \\
    \midrule
    Emphysema & 0.83    & \textbf{0.94}  & 0.92 & \textbf{0.94} & 0.91  \\
    \midrule
    Fibrosis & 0.79    & 0.83  & 0.83 & 0.82 & \textbf{0.84}  \\
    \midrule
    Pleural\_Thickening & 0.68   & 0.79  & 0.79 & 0.81 & \textbf{0.89} \\
    \midrule
    Hernia & 0.87  & 0.91  & \textbf{0.94} & 0.92 & 0.90  \\
    \midrule
    Mean AUC & 0.745  & 0.819  & 0.832 &0.850 & \textbf{0.867} \\
    \bottomrule
    \end{tabular}%
  \label{tb1}%
\end{table}%

From these results, we have the following observations:
\begin{itemize}
    \item In terms of the averaged AUC score, our proposed method performs best, our method achieved results comparable to the optimal algorithms for diagnosing most pathologies. 
    
    \item Taking ``pneumonia'' for example, this pathology is relatively uncommon, comprising less than 1\% of the total dataset. Due to their rarity, all the compared methods performed poorly in detecting it, likely due to the long tail distribution of the data. Besides, the co-occurrence analysis in \cite{wang2017chestx} also shows that pneumonia is easy to misdiagnose, indicating that the classification of pneumonia is a challenging task. In contrast, our method leverages instrumental variables to guide the causal feature learning for the final pathology representation. 

    \item With the semantic IV, compared with other previous work, we can find our method is more explainable as shown in Fig.\ref{example}, where our method takes more attention to the accurate pathological areas. In a word, we've used the IV learning part to find the true cause of the disease, whereas other comparable methods only rely on the quality of training data to capture the relationship between CXR images. 
\end{itemize}

\subsubsection{CheXpert}
In this section, we evaluate the performance of our proposed framework on the CheXpert data set. We use the baseline work \cite{irvin2019chexpert} as a reference and map the uncertain labels to 1 since this has shown to work well. In addition, we evaluate five classes, known as competition tasks \cite{irvin2019chexpert}, which were selected based on clinical importance and prevalence. We report our results in Table.\ref{X}, our proposed framework achieves an average AUC score of 0.923 over five pathologies, we have the following conclusions and discussions:
\begin{itemize}
    \item We achieve the best mean AUC compared with other baselines, and get the best score of four diseases among the five challenging tasks, which proves the performance of our framework.

    \item We also explore the impact of setting the uncertain labels to ``ones'' or ``zeros'' on the model's performance. When using ``ones,'' we achieve a mean AUC of 0.923, while when using ``zeros'', we obtain a mean AUC score of 0.914, which also surpasses the corresponding baseline ``U-Zeros''. While more reasonable uncertain items filling methods and approaches could be explored to achieve better performance, this may be beyond the scope of this paper.

    \item We notice that our proposed model shows superiority in some pathologies except ``Edema.'' The performance of ``Cardiomegaly'' and ``Atelectasis'' is significantly improved by 6.85\% and 7.34\%, respectively. However, the performance of ``Pleural Effusion'' is not improved too much compared to other methods. 

\end{itemize}

\begin{table}[t]
\renewcommand\arraystretch{0.3}
\setlength\tabcolsep{4pt}%调列距
  \centering
  \caption{Comparison of AUC Scores with Different Models on CheXpert Validation Set}
    \begin{tabular}{c|c|c|c|c|c}
    \toprule
    \tabincell{c}{Abnormality}&\tabincell{c}{U-Zeros\\\cite{irvin2019chexpert}}&\tabincell{c}{U-Ones\\\cite{irvin2019chexpert}}&\tabincell{c}{CT+LSR\\\cite{pham2021interpreting}}&\tabincell{c}{Xi \textit{et al.}\\\cite{ouyang2020learning}}&\tabincell{c}{Ours}\\
    \midrule
    % \midrule
    Atelectasis   & 0.811  & 0.858  & 0.825 &0.920 & \textbf{0.921} \\
    \midrule
    Cardiomegaly   & 0.840  & 0.832  & 0.855 &0.886 & \textbf{0.889} \\
    \midrule
    Consolidation  & 0.932  & 0.899  & 0.937 &0.907 & \textbf{0.939}  \\
    \midrule
    Edema   & 0.929  & \textbf{0.941} & 0.930 &0.937 & 0.932  \\
    \midrule
    Pleural Effusion   & 0.931  & \textbf{0.934}  & 0.923 &0.933 & \textbf{0.934} \\
    \midrule
    % \midrule
    Mean AUC  & 0.889 & 0.893 & 0.894 &0.917 & \textbf{0.923} \\
    \bottomrule
    \end{tabular}%
  \label{X}%
\end{table}%

\subsubsection{MIMIC-CXR}
The AUC comparison results between our method and other works are shown in Table.\ref{tb3}, where ``Enlarged-C'' denotes ``Enlarged Cardiomediastinum''. We have the following conclusions and discussions:

\begin{itemize}
    \item We achieve the best mean AUC compared with other baselines, and we get the best AUC score for seven diseases, which proves the performance of our framework.

    \item Thanks to the abundant EHRs in the MIMIC-CXR data set, we can use it as auxiliary information to learn the better representation of the instrumental variables, which can be utilized to eliminate the confounding effect, this viewpoint is also proved in the ablation study section.

    \item We notice that in some cases like ``Pneumothorax'', our proposed model has a certain gap compared with \cite{zhu2021mvc}, we think this may be because they utilize multi-view CXR image at the same time, where we take every image as independent. We will consider this problem in future work.

\end{itemize}

\begin{table}[!ht]
\setlength\tabcolsep{3pt}
  \centering
  \caption{Comparation Of AUC Scores with previous SOTA Works on MIMIC-CXR dataset}
    \begin{tabular}{c|c|c|c|c}
    \toprule
    \tabincell{c}{Abnormality}&\tabincell{c}{MVC-NET\\\cite{zhu2021mvc}}&\tabincell{c}{ImageGCN\\\cite{mao2022imagegcn}}&\tabincell{c}{Haque \textit{et al.}\\\cite{haque2021effect}}&\tabincell{c}{Ours}\\
    \midrule
    Atelectasis & 0.818  & 0.758 & 0.808 & \textbf{0.823} \\
    \midrule
    Cardiomegaly & \textbf{0.848}  & 0.768 &0.816 & 0.825 \\
    \midrule
    Consolidation & 0.829  & 0.761 &0.821 & \textbf{0.835} \\
    \midrule
    Edema   & \textbf{0.919}  & 0.887 & 0.889 & 0.910 \\
    \midrule
    Enlarged-C   & 0.725  & 0.576 & 0.739 & \textbf{0.776} \\
    \midrule
    Fracture  & 0.665  & 0.641 & \textbf{0.667} & 0.662  \\
    \midrule
    Lung Lesion    & 0.740  & 0.673 & 0.738& \textbf{0.746}  \\
    \midrule
    Lung Opacity    & 0.757  &0.721  &0.749 & \textbf{0.767} \\
    \midrule
    Pleural Effusion    & \textbf{0.947}  & 0.904 & 0.919 & 0.913 \\
    \midrule
    Pleural Other    & 0.825  & 0.743 &0.806 & \textbf{0.831} \\
    \midrule
    Pneumonia    & 0.715  & 0.683 & 0.714 & \textbf{0.722}  \\
    \midrule
    Pneumothorax   & \textbf{0.899}  & 0.785 & 0.856 & 0.859  \\
    \midrule
    Support Devices  & -  & 0.842 & \textbf{0.898} & 0.891 \\
    \midrule
    Mean AUC  & 0.807  & 0.746 &0.805 & \textbf{0.812} \\
    \bottomrule
    \end{tabular}%
  \label{tb3}%
\end{table}%

\subsection{Ablation Study}
In this section, we conduct the ablation study to prove the superiority of our proposed thorax disease classification framework. We run the ablation experiment on the MIMIC-CXR data set to demonstrate the significance of a valid IV. The related experiments are shown in Table.\ref{abl}. ``+'' represents the framework using that element, while ``-'' means the framework removing the element. Note that ``-'' for IV learning, we train two ResNet to learn confounders and IV respectively; ``-'' for the semantic fusion module, we just concatenate the features from CXR and EHR.

\begin{table}[htbp]
\setlength\tabcolsep{4pt}%调列距
  \centering
  \caption{Ablation Study}
    \begin{tabular}{c|c|c|c|cr}
% \cmidrule{1-5}    Model & Causal Learning& Semantic Fusion& &AUC\\
\cmidrule{1-5}    Model &\tabincell{c}{IV\\Learning}&\tabincell{c}{Semantic\\Fusion}&\tabincell{c}{Valid IV\\Constraints}& AUC  \\
\cmidrule{1-5}   
    1     & -     & -   & -      & 0.717     &  \\
    2     & -     & +    & -     & 0.746     &  \\
    3     & +     & -     & -    & 0.751     &  \\
    4     & +     & +     & -    & 0.812    &  \\
    5     & -     & -   & +      & 0.725     &  \\
    6     & -     & +    & +     & 0.769     &  \\
    7     & +     & -     & +    & 0.773     &  \\
    8     & +     & +     & +    & \textbf{0.812}     &  \\
\cmidrule{1-5}    \end{tabular}%
  \label{abl}%
\end{table}%

Based on this setting, we can achieve the following observations:
\begin{itemize}
    \item The IV learning module which is designed to learn the confounders $C$ and instrumental variables $I$ has the decoupling capability according to models 1 and 3 in Table.\ref{abl}. With this module, the performance increases by about 5\%, which is a large progress.
    \item The semantic fusion module is designed to learn a semantic IV, it makes the classification framework more explainable, the corresponding visualization results are shown in Fig.\ref{example}, where the model pays more attention to the true cause of the disease. Model 1 and 2 proves the rationality of the semantic and fusion module.
    \item The framework with IV learning and semantic fusion modules, i.e. model 4 performs well compared with models 2 and 3 with an only module, which indicates the whole SCM setting of our task is relatively complete and reasonable, all components are organically integrated into one overall framework.
    \item The valid IV constraints of our framework, i.e. the proposed loss functions in Section.\ref{constarints}, successfully constraints the mutual information between causal variables, we can find when the constraints are applied, the corresponding classification performance has been improved compared with the model when we remove the related settings.
\end{itemize}

In general, our proposed framework includes all modules, which gives us the best mean AUC of 0.812 on the MIMIC-CXR dataset. The corresponding experiments and analyses demonstrate the performance of these modules.

\section{CONCLUSIONS AND DISCUSSION}
\label{sec:conclution}
In this paper, we propose an instrumental variable learning framework to solve the multi-label CXR image classification problem.
We first design a SCM for the task and apply the deep model to learn the related variables. Specifically, we adopt an instrumental variable method to eliminate the confounding effect and learn the IV from the CXR data and auxiliary information like EHRs with some constraints to make IV valid.
We evaluated our proposed method on three public data sets, NIH ChestX-ray14, CheXpert, and MIMIC-CXR. Standard experimental results indicate that our proposed framework can achieve state-of-the-art results on the thorax abnormal diseases classification task. 

We next briefly discuss the limitations of our proposed method and future work possibilities. We have some assumptions such as we assume that a valid IV can be fully learned from CXR data and related EHRs, we may relax such assumptions and explore more reasonable methods. We will also try to prove the robustness and applicability of our framework in other medical problem scenarios or other kinds of problems, i.e. image segmentation.




% \section*{ACKNOWLEDGMENT}

% Put sponsor acknowledgments in the unnumbered footnote on the first page.

\bibliographystyle{ieeetr}
\bibliography{mybib}

\begin{thebibliography}{10}

\bibitem{brady2012discrepancy}
A.~Brady, R.~{\'O}. Laoide, P.~McCarthy, and R.~McDermott, ``Discrepancy and
  error in radiology: concepts, causes and consequences,'' {\em The Ulster
  medical journal}, vol.~81, no.~1, p.~3, 2012.

\bibitem{mao2022imagegcn}
C.~Mao, L.~Yao, and Y.~Luo, ``Imagegcn: Multi-relational image graph
  convolutional networks for disease identification with chest x-rays,'' {\em
  IEEE Transactions on Medical Imaging}, vol.~41, no.~8, pp.~1990--2003, 2022.

\bibitem{ouyang2020learning}
X.~Ouyang, S.~Karanam, Z.~Wu, T.~Chen, J.~Huo, X.~S. Zhou, Q.~Wang, and J.-Z.
  Cheng, ``Learning hierarchical attention for weakly-supervised chest x-ray
  abnormality localization and diagnosis,'' {\em IEEE Transactions on Medical
  Imaging}, 2020.

\bibitem{glymour2016causal}
M.~Glymour, J.~Pearl, and N.~P. Jewell, {\em Causal inference in statistics: A
  primer}.
\newblock John Wiley \& Sons, 2016.

\bibitem{pearl2000models}
J.~Pearl {\em et~al.}, ``Models, reasoning and inference,'' {\em Cambridge, UK:
  CambridgeUniversityPress}, vol.~19, no.~2, 2000.

\bibitem{pearl2014interpretation}
J.~Pearl, ``Interpretation and identification of causal mediation.,'' {\em
  Psychological methods}, vol.~19, no.~4, p.~459, 2014.

\bibitem{sui2022causal}
Y.~Sui, X.~Wang, J.~Wu, M.~Lin, X.~He, and T.-S. Chua, ``Causal attention for
  interpretable and generalizable graph classification,'' in {\em Proceedings
  of the 28th ACM SIGKDD Conference on Knowledge Discovery and Data Mining},
  pp.~1696--1705, 2022.

\bibitem{ribeiro2023learning}
M.~d.~C. Ribeiro-Dantas, H.~Li, V.~Cabeli, L.~Dupuis, F.~Simon, L.~Hettal,
  A.-S. Hamy, and H.~Isambert, ``Learning interpretable causal networks from
  very large datasets, application to 400,000 medical records of breast cancer
  patients,'' {\em arXiv preprint arXiv:2303.06423}, 2023.

\bibitem{pearl2009causal}
J.~Pearl, ``Causal inference in statistics: An overview,'' {\em Statistics
  surveys}, 2009.

\bibitem{rajaraman2020training}
S.~Rajaraman and S.~Antani, ``Training deep learning algorithms with weakly
  labeled pneumonia chest x-ray data for covid-19 detection,'' {\em MedRxiv},
  2020.

\bibitem{wang2017chestx}
X.~Wang, Y.~Peng, L.~Lu, Z.~Lu, M.~Bagheri, and R.~M. Summers, ``Chestx-ray8:
  Hospital-scale chest x-ray database and benchmarks on weakly-supervised
  classification and localization of common thorax diseases,'' in {\em
  Proceedings of the IEEE conference on computer vision and pattern
  recognition}, pp.~2097--2106, 2017.

\bibitem{zhang2021deep}
Y.~Zhang, B.~Kang, B.~Hooi, S.~Yan, and J.~Feng, ``Deep long-tailed learning: A
  survey,'' {\em arXiv preprint arXiv:2110.04596}, 2021.

\bibitem{selvaraju2017grad}
R.~R. Selvaraju, M.~Cogswell, A.~Das, R.~Vedantam, D.~Parikh, and D.~Batra,
  ``Grad-cam: Visual explanations from deep networks via gradient-based
  localization,'' in {\em Proceedings of the IEEE international conference on
  computer vision}, pp.~618--626, 2017.

\bibitem{yue2020interventional}
Z.~Yue, H.~Zhang, Q.~Sun, and X.-S. Hua, ``Interventional few-shot learning,''
  {\em Advances in neural information processing systems}, vol.~33,
  pp.~2734--2746, 2020.

\bibitem{rocha2022attention}
J.~Rocha, S.~C. Pereira, J.~Pedrosa, A.~Campilho, and A.~M. Mendon{\c{c}}a,
  ``Attention-driven spatial transformer network for abnormality detection in
  chest x-ray images,'' in {\em 2022 IEEE 35th International Symposium on
  Computer-Based Medical Systems (CBMS)}, pp.~252--257, IEEE, 2022.

\bibitem{minaee2020deep}
S.~Minaee, R.~Kafieh, M.~Sonka, S.~Yazdani, and G.~J. Soufi, ``Deep-covid:
  Predicting covid-19 from chest x-ray images using deep transfer learning,''
  {\em Medical image analysis}, vol.~65, p.~101794, 2020.

\bibitem{ke2021chextransfer}
A.~Ke, W.~Ellsworth, O.~Banerjee, A.~Y. Ng, and P.~Rajpurkar, ``Chextransfer:
  performance and parameter efficiency of imagenet models for chest x-ray
  interpretation,'' in {\em Proceedings of the Conference on Health, Inference,
  and Learning}, pp.~116--124, 2021.

\bibitem{paul2021discriminative}
A.~Paul, Y.-X. Tang, T.~C. Shen, and R.~M. Summers, ``Discriminative ensemble
  learning for few-shot chest x-ray diagnosis,'' {\em Medical image analysis},
  vol.~68, p.~101911, 2021.

\bibitem{he2016deep}
K.~He, X.~Zhang, S.~Ren, and J.~Sun, ``Deep residual learning for image
  recognition,'' in {\em Proceedings of the IEEE conference on computer vision
  and pattern recognition}, pp.~770--778, 2016.

\bibitem{huang2017densely}
G.~Huang, Z.~Liu, L.~Van Der~Maaten, and K.~Q. Weinberger, ``Densely connected
  convolutional networks,'' in {\em Proceedings of the IEEE conference on
  computer vision and pattern recognition}, pp.~4700--4708, 2017.

\bibitem{ratul2021multi}
R.~H. Ratul, F.~A. Husain, T.~H. Purnata, R.~A. Pomil, S.~Khandoker, and M.~Z.
  Parvez, ``Multi-stage optimization of deep learning model to detect thoracic
  complications,'' in {\em 2021 IEEE International Conference on Systems, Man,
  and Cybernetics (SMC)}, pp.~3000--3005, IEEE, 2021.

\bibitem{sobel1996introduction}
M.~E. Sobel, ``An introduction to causal inference,'' {\em Sociological Methods
  \& Research}, vol.~24, no.~3, pp.~353--379, 1996.

\bibitem{richiardi2013mediation}
L.~Richiardi, R.~Bellocco, and D.~Zugna, ``Mediation analysis in epidemiology:
  methods, interpretation and bias,'' {\em International journal of
  epidemiology}, vol.~42, no.~5, pp.~1511--1519, 2013.

\bibitem{bareinboim2012controlling}
E.~Bareinboim and J.~Pearl, ``Controlling selection bias in causal inference,''
  in {\em Artificial Intelligence and Statistics}, pp.~100--108, PMLR, 2012.

\bibitem{wang2020visual}
T.~Wang, J.~Huang, H.~Zhang, and Q.~Sun, ``Visual commonsense r-cnn,'' in {\em
  Proceedings of the IEEE/CVF Conference on Computer Vision and Pattern
  Recognition}, pp.~10760--10770, 2020.

\bibitem{yang2021deconfounded}
X.~Yang, H.~Zhang, and J.~Cai, ``Deconfounded image captioning: A causal
  retrospect,'' {\em IEEE Transactions on Pattern Analysis and Machine
  Intelligence}, 2021.

\bibitem{zhang2020causal}
D.~Zhang, H.~Zhang, J.~Tang, X.-S. Hua, and Q.~Sun, ``Causal intervention for
  weakly-supervised semantic segmentation,'' {\em Advances in Neural
  Information Processing Systems}, vol.~33, pp.~655--666, 2020.

\bibitem{li2021causal}
J.~Li, B.~Wu, X.~Sun, and Y.~Wang, ``Causal hidden markov model for time series
  disease forecasting,'' in {\em Proceedings of the IEEE/CVF Conference on
  Computer Vision and Pattern Recognition}, pp.~12105--12114, 2021.

\bibitem{yuan2022auto}
J.~Yuan, A.~Wu, K.~Kuang, B.~Li, R.~Wu, F.~Wu, and L.~Lin, ``Auto iv:
  Counterfactual prediction via automatic instrumental variable
  decomposition,'' {\em ACM Transactions on Knowledge Discovery from Data
  (TKDD)}, vol.~16, no.~4, pp.~1--20, 2022.

\bibitem{vaswani2017attention}
A.~Vaswani, N.~Shazeer, N.~Parmar, J.~Uszkoreit, L.~Jones, A.~N. Gomez,
  {\L}.~Kaiser, and I.~Polosukhin, ``Attention is all you need,'' in {\em
  Advances in neural information processing systems}, pp.~5998--6008, 2017.

\bibitem{huang2022clinicalbert}
K.~Huang, J.~Altosaar, and R.~Ranganath, ``Clinicalbert: modeling clinical
  notes and predicting hospital readmission. arxiv: 190405342 [cs] published
  online first: 28 november 2020,'' 2022.

\bibitem{irvin2019chexpert}
J.~Irvin, P.~Rajpurkar, M.~Ko, Y.~Yu, S.~Ciurea-Ilcus, C.~Chute, H.~Marklund,
  B.~Haghgoo, R.~Ball, K.~Shpanskaya, {\em et~al.}, ``Chexpert: A large chest
  radiograph dataset with uncertainty labels and expert comparison,'' in {\em
  Proceedings of the AAAI conference on artificial intelligence}, vol.~33,
  pp.~590--597, 2019.

\bibitem{johnson2019mimic}
A.~E. Johnson, T.~J. Pollard, S.~J. Berkowitz, N.~R. Greenbaum, M.~P. Lungren,
  C.-y. Deng, R.~G. Mark, and S.~Horng, ``Mimic-cxr, a de-identified publicly
  available database of chest radiographs with free-text reports,'' {\em
  Scientific data}, vol.~6, no.~1, p.~317, 2019.

\bibitem{kingma2014adam}
D.~P. Kingma and J.~Ba, ``Adam: A method for stochastic optimization,'' {\em
  arXiv preprint arXiv:1412.6980}, 2014.

\bibitem{loshchilov2017decoupled}
I.~Loshchilov and F.~Hutter, ``Decoupled weight decay regularization,'' {\em
  arXiv preprint arXiv:1711.05101}, 2017.

\bibitem{gong2021deformable}
X.~Gong, X.~Xia, W.~Zhu, B.~Zhang, D.~Doermann, and L.~Zhuo, ``Deformable gabor
  feature networks for biomedical image classification,'' in {\em Proceedings
  of the IEEE/CVF Winter Conference on Applications of Computer Vision},
  pp.~4004--4012, 2021.

\bibitem{pham2021interpreting}
H.~H. Pham, T.~T. Le, D.~Q. Tran, D.~T. Ngo, and H.~Q. Nguyen, ``Interpreting
  chest x-rays via cnns that exploit hierarchical disease dependencies and
  uncertainty labels,'' {\em Neurocomputing}, vol.~437, pp.~186--194, 2021.

\bibitem{zhu2021mvc}
X.~Zhu and Q.~Feng, ``Mvc-net: Multi-view chest radiograph classification
  network with deep fusion,'' in {\em 2021 IEEE 18th International Symposium on
  Biomedical Imaging (ISBI)}, pp.~554--558, IEEE, 2021.

\bibitem{haque2021effect}
M.~I.~U. Haque, A.~K. Dubey, and J.~D. Hinkle, ``The effect of image resolution
  on automated classification of chest x-rays,'' {\em Medrxiv}, pp.~2021--07,
  2021.

\end{thebibliography}

\end{document}